\documentclass{article}
\usepackage{spconf,amsmath,graphicx}
\usepackage{graphicx}
\usepackage[colorlinks=true, allcolors=blue]{hyperref}
\usepackage{outlines}
\usepackage{booktabs}
\usepackage{algorithm}
\usepackage[noend]{algpseudocode}
\usepackage{multirow}
\usepackage{subcaption}
\usepackage[normalem]{ulem}
\usepackage{tikz}
\useunder{\uline}{\ul}{}

\title{Subjective assessment of the impact of a content adaptive optimiser for compressing 4K HDR content with AV1}
\name{Vibhoothi, Angeliki Katsenou, François Pitié, Katarina Domijan, Anil Kokaram \thanks{This work was funded by DTIF EI Grant No DT-2019-0068 and The ADAPT SFI Research Center.}}
\address{Sigmedia Group,\\
Department of Electronic and Electrical Engineering, \\
Trinity College Dublin, Ireland. \\
\{vibhoothi, akatsenou, pitief, anil.kokaram\}@tcd.ie}
\name{Vibhoothi$^{\dagger}$, Angeliki Katsenou$^{\dagger}$, François Pitié$^{\dagger}$, Katarina Domijan$^{\ddagger}$, Anil Kokaram$^{\dagger}$\thanks{This work was funded by DTIF EI Grant No DT-2019-0068 and The ADAPT SFI Research Center.}}
\address{$^{\dagger}$ Sigmedia Group,
Department of Electronic and Electrical Engineering, 
Trinity College Dublin, Ireland. \\$^{\ddagger}$Department of Mathematics and Statistics, Maynooth University, Ireland\\$^{\dagger}$
\{vibhoothi, akatsenou, pitief, anil.kokaram\}@tcd.ie
$^{\ddagger}$\{katarina.domijan\}@mu.ie}
\begin{document}
\ninept

\maketitle
\begin{abstract}
Since 2015 video dimensionality has expanded to higher spatial and temporal resolutions and a wider colour gamut. This High Dynamic Range (HDR) content has gained traction in the consumer space as it delivers an enhanced quality of experience. At the same time, the complexity of codecs is growing. This has driven the development of tools for content-adaptive optimisation that achieve optimal rate-distortion performance for HDR video at 4K resolution.
While improvements of just a few percentage points in BD-Rate (1-5\%) are significant for the streaming media industry, the impact on subjective quality has been less studied especially for HDR/AV1. 
In this paper, we conduct a subjective quality assessment (42 subjects) of 4K HDR content with a per-clip optimisation strategy.  We correlate these subjective scores with existing popular objective metrics used in standard development and show that some perceptual metrics correlate surprisingly well even though they are not tuned for HDR.  We find that the DSQCS protocol is too insensitive to categorically compare the methods but the data allows us to make recommendations about the use of experts vs non-experts in HDR studies, and explain the subjective impact of film grain in HDR content under compression. 


\end{abstract}
\begin{keywords}
HDR, AV1, Rate-Distortion optimisation, Subjective study, Quality metrics.
\end{keywords}

\section{Introduction}
\label{sec:intro}

High Dynamic Range (HDR) content distribution has followed the increasing availability and reducing cost of HDR displays since 2015~\cite{2016dufauxhdrbook,2022smpteuhdprogress}. As part of the UHD-4K sub-genre, HDR content, distributed at 10bits is typical of notoriously high bitrate. Per-clip compression strategies \cite{2019netflixPerClip}, championed by Netflix since 2015 are therefore even more important at this high data rate.  

 Work targetting HDR for AVC and HEVC encoders~\cite{2014lambda_hdr_avc_subj} and ~\cite{2019hdrhevcsubjective, 2022hdrambient} respectively, highlights that the subjective impact of changes to HDR encoding is quantitatively different from standard dynamic range (SDR) content. In our previous work~\cite{vibspie2022}, we demonstrated that the Lagrange multiplier ($\lambda$) can be optimised on a per-clip basis using BD-Rate as a cost function within a classical numerical optimisation scheme. We observed gains in BD-Rate (\%) of MS-SSIM of 1.63\% in AV1 on average for 4K HDR sequences. These small gains become significant at scale but there is little work on subjective studies to validate objective-metric gains like these from an actually-perceived point of view, and even more so for AV1. 
 
Our main contributions are therefore: i) subjective evaluation of the numerical optimiser for the $\lambda$ multiplier using seven 4K HDR videos for AV1, ii) a statistical analysis of results regarding expert VS non-experts in HDR studies iii) evidence for the correlation of subjective scores with current HDR metrics and the impact of film grain in HDR quality. 
We are also releasing our dataset and MOS scores~\cite{icip2023-suppl} to support HDR video quality metric development by the research community. 

Section~\ref{sec:background} presents the background and previous subjective studies for HDR videos. Section~\ref{sec:sub-expt} shows the experimental setup, and Section~\ref{sec:result-discsn} reports on results and analysis. 

\section{Background}
\label{sec:background}

\subsection{Per-clip $\lambda$ Optimisation}
\label{sec:background:rdo}
Rate-Distortion Optimisation (RDO) is a delicate constrained minimisation problem inside the current modern hybrid video encoders. 
The main idea of the RDO is to find the right set of parameters to achieve the lowest visual distortion ($D$) at a target bitrate ($R$). In order to deal with the increasing number of parameters inside the encoders, Sullivan et. al~\cite{sullivan1998rate} proposed to cast this constrained optimisation into a more manageable unconstrained optimisation, through the use of a combined RD tradeoff, i.e. $J = D + \lambda R$, defined by a Lagrange multiplier, $\lambda$. The minimisation of $J$ for any $\lambda$ yields a Pareto optimal pair $(R, D)$. The problem is to find the value of $\lambda$ that results in the desired target rate $R$.
Different video encoders devised different recipes to derive the optimal value of $\lambda$ from $qp$, the quantiser step size, which is an impactful parameter in compression. Increasing $qp$ reduces the rate $R$ but also increases distortion $D$. 

In the libaom-AV1 codebase, $\lambda$ is estimated as $\lambda \approx A \times q_{dc} ^2$, where $A\in [3.2, 4.2]$ is a constant based on the frame type, and $q_{dc}=f(q_i, A)$ is a discrete-valued lookup table. Quantizer index, $q_i \in [0..255]$, is approximately $\approx 4 \times qp$ ($qp \in [0..63]$). 
This $\lambda\mbox{-}qp$ relationship is not universally optimal~\cite{icip2022paper}. It is rather an empirical relationship that was derived with experimentation over a set of test videos typically used in standards. Thus, to maximise gains, $\lambda$ can be customised per content.

The idea of $\lambda$ adaptation based on video content is not new. In 2015, Zhang and Bull~\cite{zhangbulllambdahevcv2} altered $\lambda$ based on distortion statistics on a frame-basis for HEVC. In 2020, Ringis et. al~\cite{EIRingis, pcs2021ringis} established $\lambda$ tuning on a per-clip basis using a single modified $\lambda = k\cdot\lambda_0$ ($k$ is $\lambda$ multiplier, $\lambda_0$ is default encoder choice $\lambda$) across all the frames can yield average BD-Rate gains of 1.63\% for VP9 (libvpx-vp9) and 1.87\% for HEVC (x265) on a large corpus of 10k videos. 
In 2022~\cite{icip2022paper}, we extended this approach and found that tuning $\lambda$ for specific frame-types can give significant improvement for AV1 and HEVC, with average BD-Rate gains of 4.92\% from 0.54\%. In~\cite{vibspie2022} we further demonstrated that this can be extended to achieve improvements for 4K HDR content.

In this work, we build the optimisation framework based on the methodology established previously~\cite{vibspie2022, icip2022paper}, which uses a numeric optimiser minimising a $\lambda$ multiplier based on Powell's Method. 
We use a $\lambda$ multiplier $k$ as $\lambda = k\cdot\lambda_0$ where $\lambda_0$ is default encoder $\lambda$. Based on the latest study~\cite{vibspie2022}, we adopted the use of two $k$s, $k_1$ for Keyframes (KF) and $k_2$ for Golden frames/Alternate-reference frames (GF/ARF). The $\lambda$s are adjusted per clip. The adjustment is guided by the BD-Rate (\%) gains on Rate-MS-SSIM curves. 
The cost function for optimisation is BD-Rate~\cite{aomctc, bdrate_extended} over the default encoder configuration at $(k1, k2) = (1,1)$. 

\vspace{-0.5em}
\subsection{HDR Subjective Testing}

\label{sec:background:subtest}
Since the standardization of HDR Perceptual Quantizer (PQ) for broadcast in ITU BT.2100~\cite{itu_hdr},  HDR displays and HDR content delivery has risen in importance~\cite{2022smpteuhdprogress}. However, there are limited subjective studies for HDR video content.
In 2014, Azimi et. al~\cite{2014lambda_hdr_avc_subj} introduced a fine-tuned $\lambda$ multiplier for H.264 based on the HDR-VDP2~\cite{hdrvdp2} metric for four videos. Results show that an average improvement of 37\% in bitrate and 15 points in the MOS scale was reported from 18 subjects. 
In 2019, Athar et. al~\cite{2019hdrhevcsubjective} conducted a detailed subjective study with 51 subjects of HDR-WCG compression with H.264 and HEVC with 4K HDR videos (14) showing that  full reference SDR objective metrics have a high correlation with subjective scores and can be utilised for HDR-WCG content evaluation.
%
%
Recently, Shang et. al.~\cite{2022hdrambient, 2023hdrsportshevc} conducted a subjective evaluation of HDR video content with HEVC first in different ambient lighting conditions for generic content and  sports content. In the first work, they found no statistically significant difference in opinion scores with dark or ambient lighting, and in the latter, they quantified the deviation of subjective preference with objective metrics. 
All these studies were performed on H.264 or HEVC-encoded content.

For AV1, there has only been a limited amount of subjective studies. 
In 2018 Akyazi et. al.~\cite{2018av1subjectivetouradj} conducted a subjective study to assess the compression efficiency of AV1 over HEVC and VP9. 
In 2019, Katsenou et. al~\cite{2019av1subjective-angkats} demonstrated that there is no significant difference between AV1 and HEVC.
In both of these studies, the focus was the codec comparison of SDR video content.

\begin{table*}[h]
\resizebox{\textwidth}{!}{%
\begin{tabular}{@{}lcccccccccccc@{}}
\toprule
\textbf{Clip}  & \textbf{PSNR-Y}  & \textbf{MS-SSIM} & \textbf{CIEDE2000} & \textbf{VMAF} & \textbf{DE100} & \textbf{L100} & \textbf{HDR-VQM} & \textbf{HDR-VDP-3} & \textbf{wPSNR-Y} & \textbf{wPSNR-U} & \textbf{wPSNR-V} & \textbf{wPSNR-Avg} \\ \midrule
\textbf{MeridianFace} & -7.583 & -7.038 & -6.087 & -10.998 & -4.518 & -7.508 & -12.470 & -8.390 & -6.912 & 0.804 & -1.203 & -3.848 \\
\textbf{NocturneRoom}  & -7.714 &  -8.137 & 4.111 & -12.653 & 12.942 & -1.423 & -5.683 & 1.349 & -7.754 & 1.752 & 48.596 & 0.913 \\
\textbf{SVTSmithyHouse}  & 3.230 & -3.942 & 2.138 & 6.088 & 1.307 & 1.128 & -4.082 & 0.179 & 5.654 & 6.863 & -2.393 & 4.663 \\
\textbf{CablesButterfly} & 0.460  & -1.232 & -3.054 & 1.404 & -4.731 & -0.771 & -7.985 & -3.072 & 0.403 & -1.925 & -5.890 & -2.556 \\
\textbf{CosmosSheep} & -6.787  & -6.348 & -2.485 & -8.298 & -8.494 & -7.515 & -8.316 & -6.812 & -6.900 & -8.573 & -8.462 & -7.897 \\
\textbf{SLFireDragon} & -5.900 & -4.999 & -6.258 & -6.787 & -5.968 & -7.204 & -9.629 & -6.490 & -6.602 & -7.495 & -6.684 & -6.815 \\
\textbf{MeridianSeaShore (T)} & -3.711 & -4.669 & -2.376 & -4.957 & 1.867 & -2.733 & 0.188 & -3.878 & -2.758 & -1.775 & 2.922 & -1.330 \\ \midrule
\textbf{Average} & \textbf{-4.001} & \textbf{-5.195} & \textbf{-2.001} & \textbf{-5.172} & \textbf{-1.085} & \textbf{-3.718} & \textbf{-6.854} & \textbf{-3.873} & \textbf{-3.553} & \textbf{-1.478} & \textbf{3.841} & \textbf{-2.410} \\ \bottomrule
\end{tabular}%
}
\caption{Objective metrics gains (\%) for the Clips, all the metrics except VMAF were represented in dB unit, negative is better. The BD-rate~\cite{bdrate_extended} is measured with 4 $qp$ points.}
\vspace{-0.5em}
\label{tab:obj-metrics}
\end{table*}

\vspace{-0.5em}
\section{Evaluation}
\label{sec:sub-expt}

We report on a double stimuli subjective evaluation of the performance of content-based optimisation for AV1 (libaom-av1) \cite{icip2022paper,vibspie2022}. The process optimises $\lambda$ for Keyframes and GF/ARF frames.

\vspace{-0.8em}
\subsection{Dataset}
\label{sec:methodology:dataset}
We use seven videos of 5-sec duration at 4K (3840$\times$2160) with BT.2020 colour primaries and SMPTE2084 Perceptual Quantizer (PQ) transfer function represented in YUV container. These are curated from Netflix open-content~\cite{netflixopencontent}, SVT open-content~\cite{svt2022}, and Cables-4K~\cite{cablelabs}, with 24 to 60 fps.
The sequences were selected based on their low-level features~\cite{itup910}; Spatial Information (SI) and Temporal Information (TI), and the BD-Rate gains~\cite{vibspie2022} for a balanced and representative dataset. For the selected long-version clips, SI is between 11.86 and 67.861 (Average 25.89) and TI is between 3.728 and 33.953 (Average 15.89). 
CosmosSheep and SLFireDragon videos are animated sequences with high amounts of motion. SVTSmithyHouse and CablesButteryfly videos are natural scenes without significant levels of acquisition noise and are almost static. SVTSmithyHouse includes many different types of static textures with medium to high granularity. NocturneRoom and MeridianFace are dark sequences, one with a person moving and one quite static but with a focus on the face. MeridianSeaShore which was used for training is a low-light sequence that depicts a seaside landscape with a person walking. These last three sequences exhibit noticeable ISO noise attributed mainly to artistic intent.
More information on the sequences can be found in the supplementary material~\cite{icip2023-suppl}.

\begin{figure}
    \includegraphics[width=.32\columnwidth]{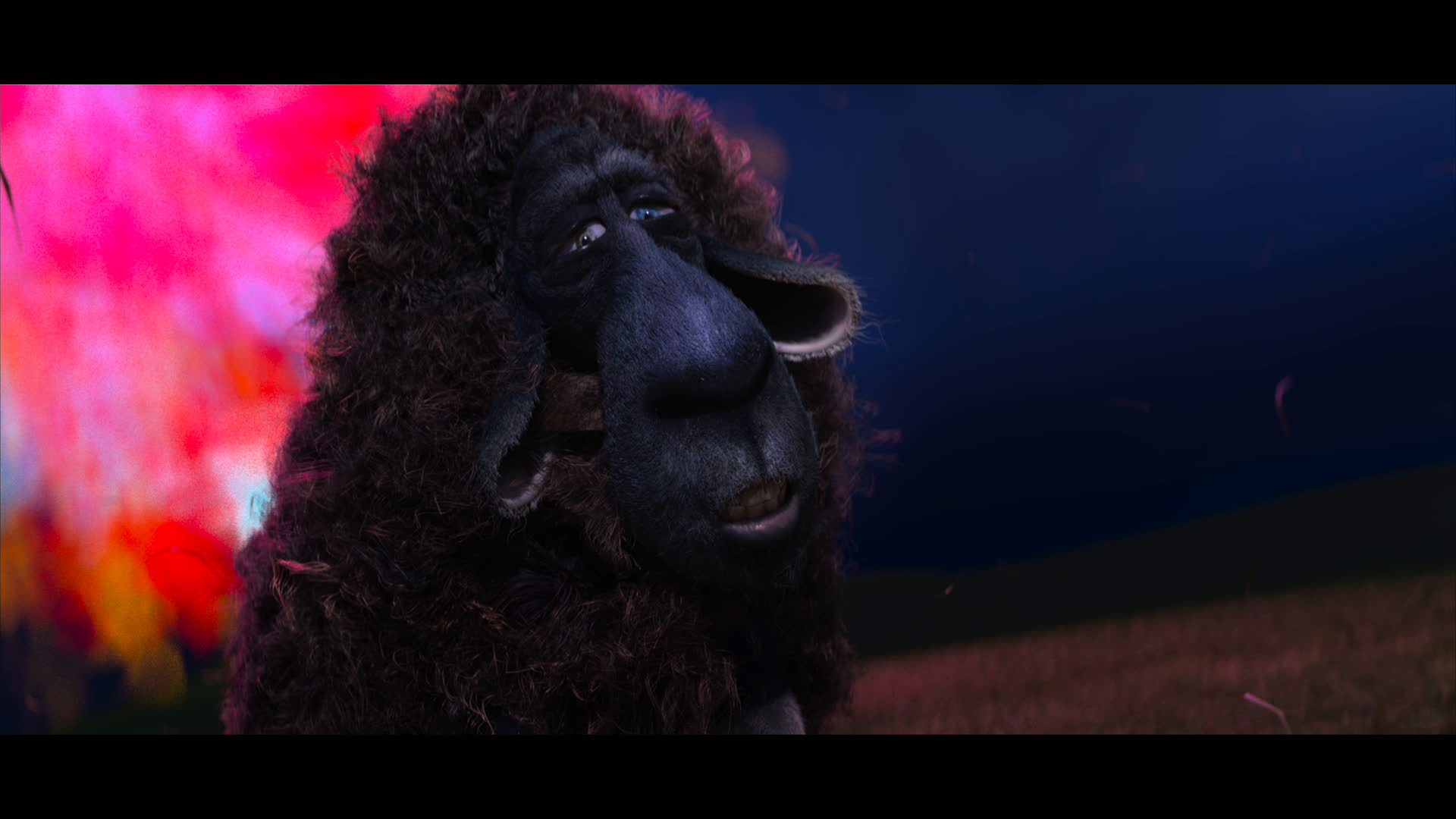}\hfill
    \includegraphics[width=.32\columnwidth]{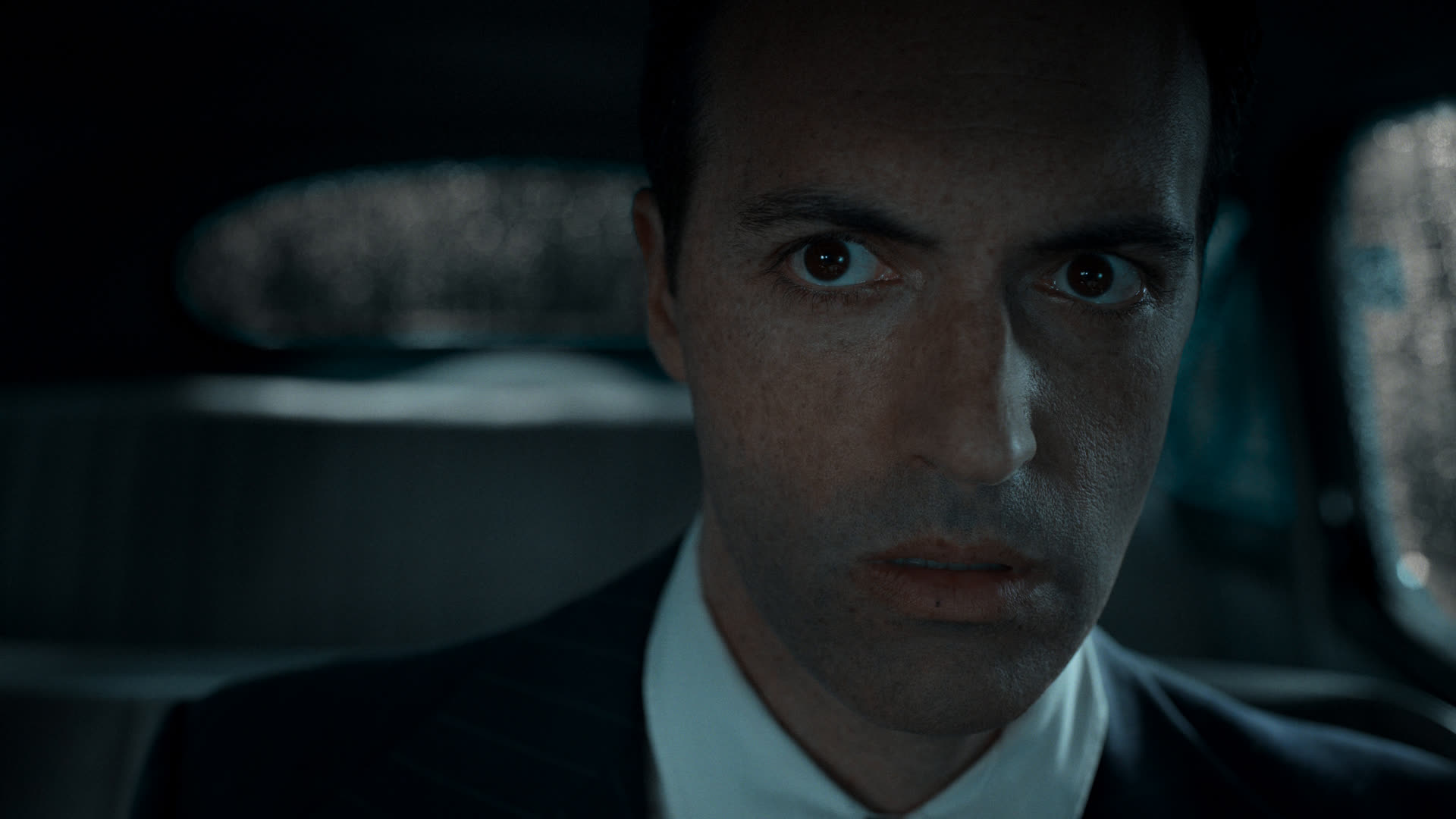}\hfill
    \includegraphics[width=.32\columnwidth]{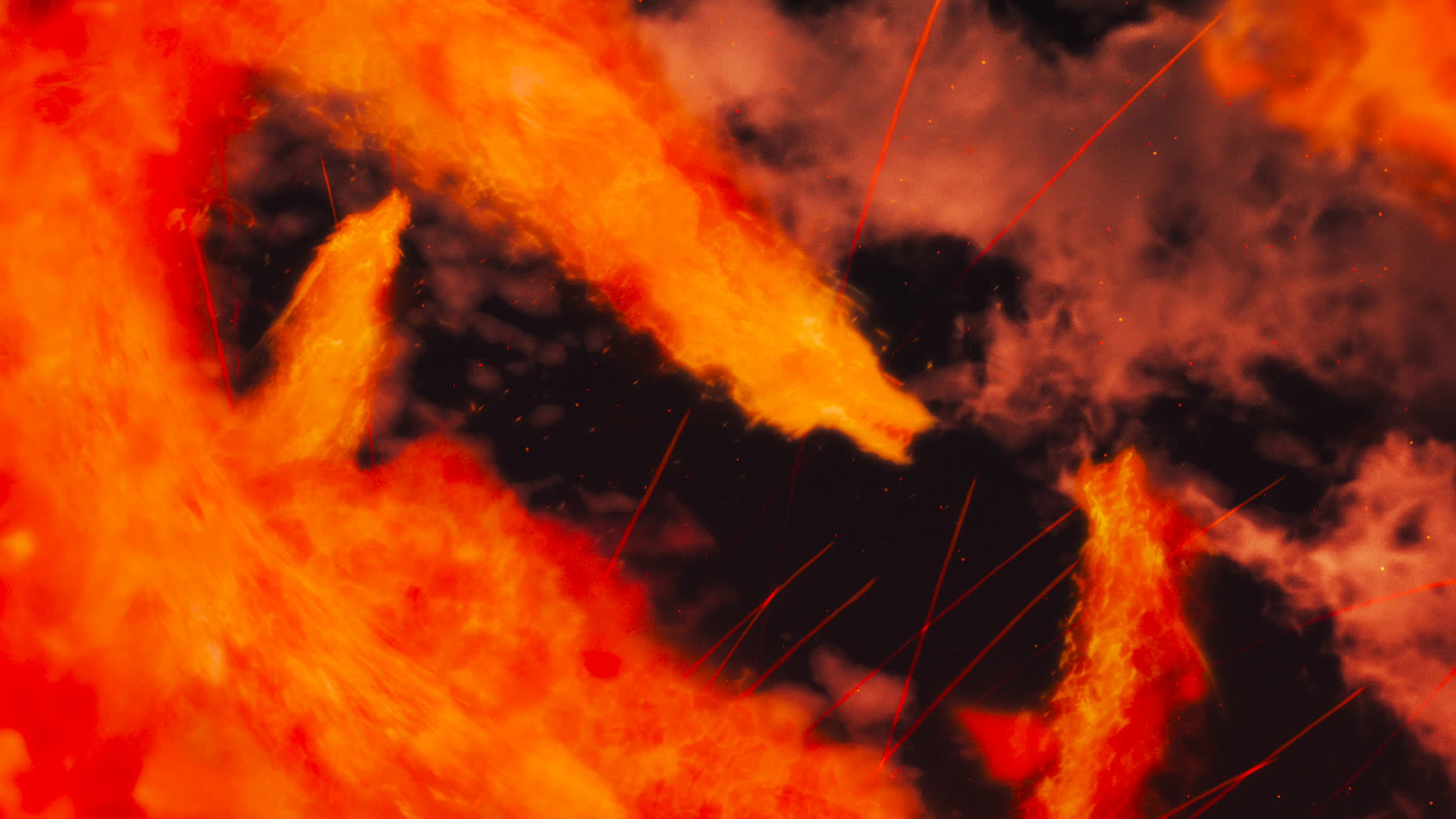}
    \\[\smallskipamount]
    \includegraphics[width=.32\columnwidth]{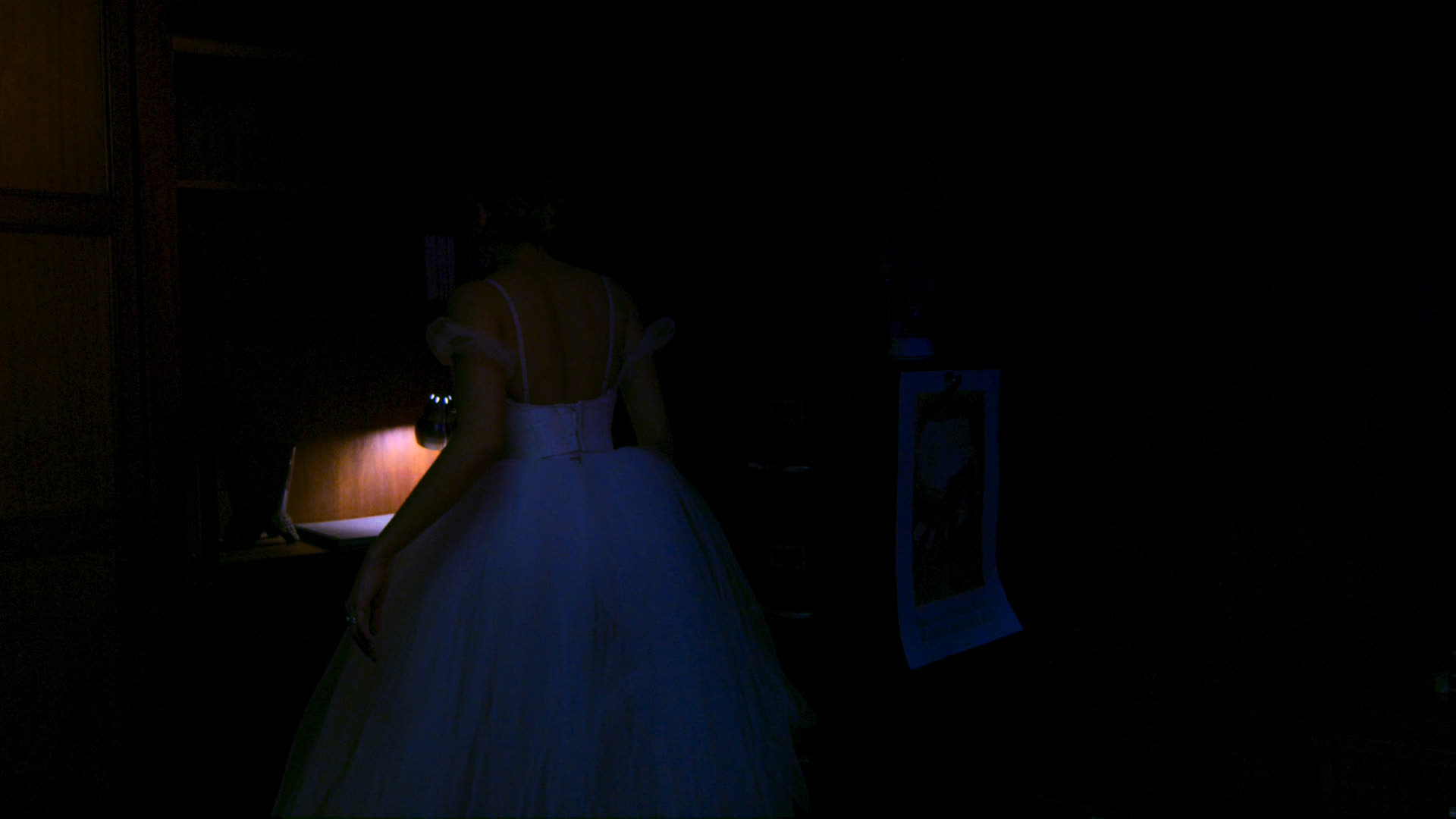}\hfill
    \includegraphics[width=.32\columnwidth]{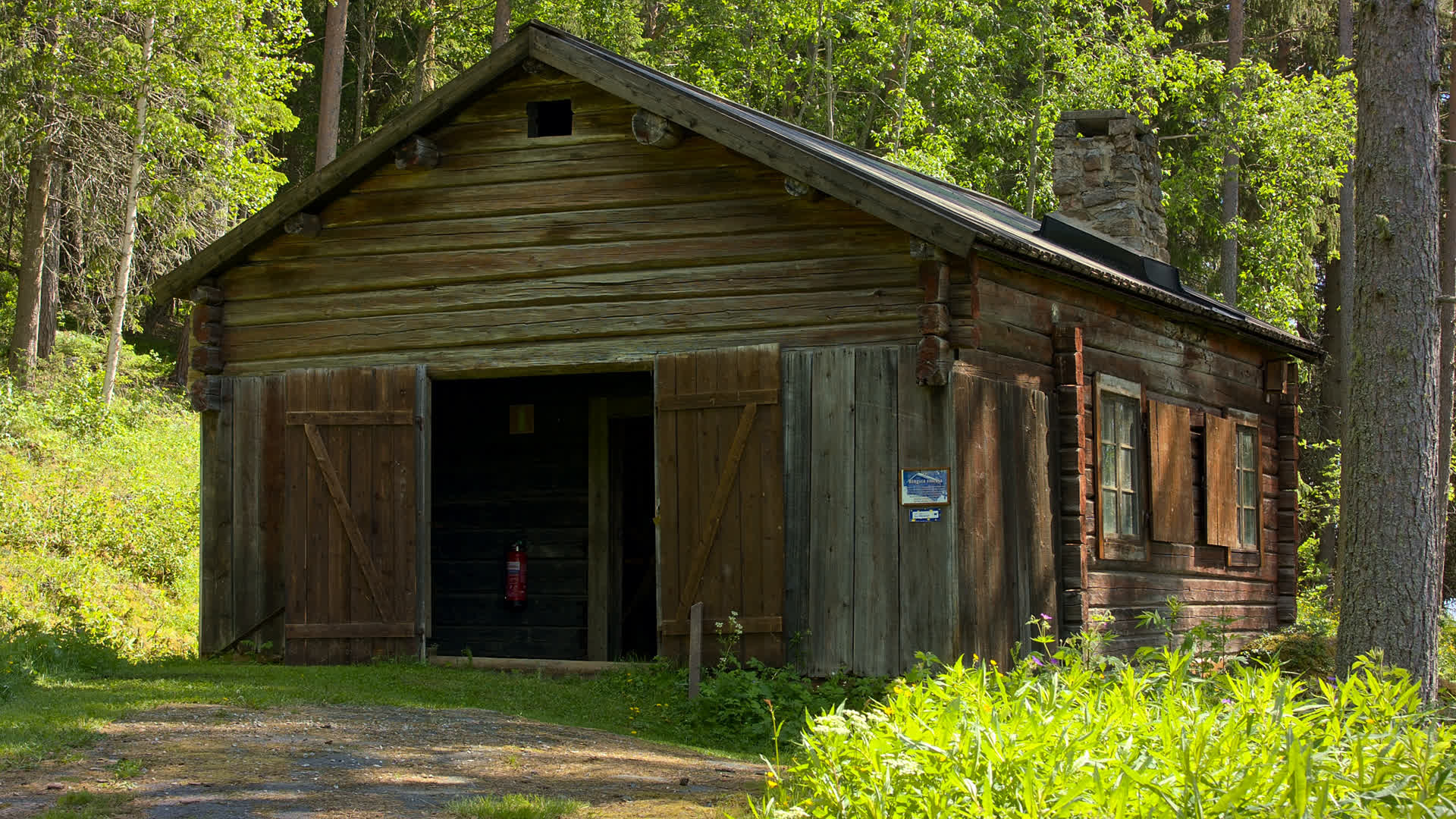}\hfill
    \includegraphics[width=.32\columnwidth]{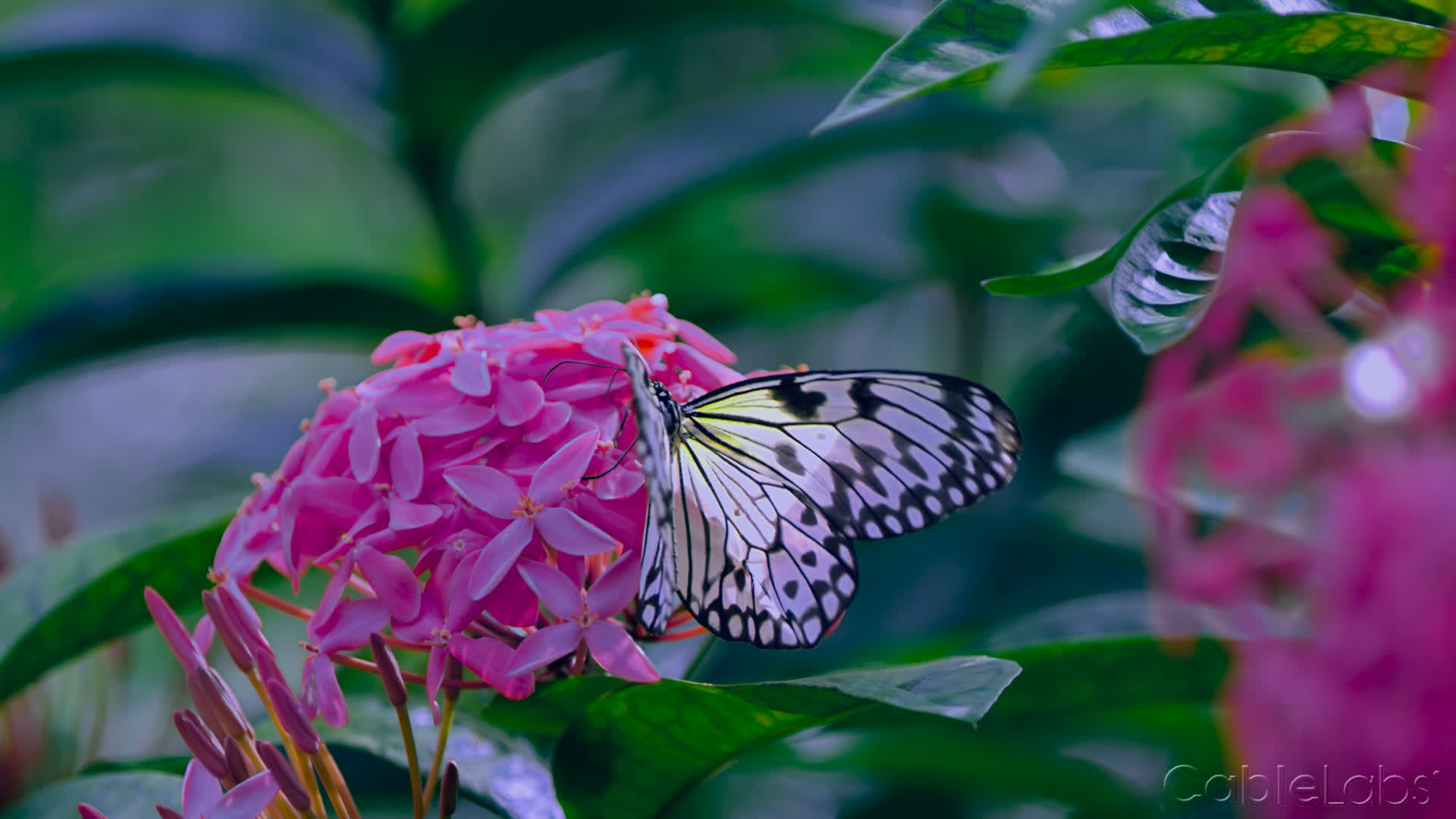}
    \vspace{-0.5em}
    \caption{Sample frames of the video dataset (Sequences are tone mapped to BT.709 for representation)}\label{fig:dataset}
    \vspace{-1em}
\end{figure}

\vspace{-0.8em}
\subsection{Environmental Setup}

\label{sec:sub-expt:env}
Subjective experiments were performed in the Sigmedia lab, Trinity College Dublin, which complies with the recommendations ITU-BT.500-14~\cite{itubt50014}. The test room uses a 32" Sony BVM-X300v2 4K/UHD OLED-critical reference monitor. Video playback used dav1d, AV1 software decoder through the Blackmagic Decklink 8K Pro connected via Quad-link 3G SDI (via FFmpeg from the workstation). HDR signal integrity was verified with JETI Spectravel 1511 Spectroradiometer. The peak brightness of the TV was measured at $\approx$1000 $cd/m^2$ (nits). The viewing distance was 1.6H. A mid-grey image 
of 14.9 nits was used during the transition between presented videos. The background luminance was controlled with a high-frequency light at 2.62 nits. The ratio of the background luminance to the picture monitor was $\approx$0.17 nits. More information on the playback and testing workflow can be found in~\cite{qomex23-vib}.

\vspace{-0.8em}
\subsection{Experimental Setup}

\label{sec:sub-expt:expt-stp}
In the optimisation, we use five operating points \sloppy
{($qp \in {27, 39, 49, 59, 63}$)} to find the optimal BD rate. On average, 66 iterations (66$\times$5=330 encodes) are needed per clip. The libaom-av1 encoder (\texttt{3.2.0, 287164d}) is deployed and the optimisation is executed at a proxy setting of a faster preset (Speed 6) along with downsampling of the source to 1920x1080 resolution (Lanczos 5). The resulting optimal $(k1, k2)$ pair is reused at native resolution with Speed 2 preset. The encoding configuration is Random Access according to AOM Common-Testing-Configuration (AOM-CTC)~\cite{aomctc}.

For this experiment, we are using a modified version of BVI-SVQA\footnote{https://github.com/goceee/BVI-SVQA, Access Feb. 2023}. For the study, we have selected 4 $qp$ points ($qp \in {27, 39, 49, 59}$) for both the default AV1 encoder and optimised versions making a total of 52 video pairs, including the ones for training. 
The protocol employed was the Double Stimulus Continuous Quality Scale (DSCQS)~\cite{itubt50014}. This protocol was selected after a small-scale expert group viewing in order to increase the confidence of the opinion scores. Hence, each pair of sequences was shown to the participant twice i.e. the original and the encoded version. Before the actual testing, a training session was conducted under the supervision of an instructor with a few representative pairs (not included in the test set). The presentation order was randomised and the participant was asked to rate videos on a [0,100] continuous scale without any time constraint. Each session had an average duration of 30mins. 

Quality analysis also included HDR objective metrics computed using HDRTools (\texttt{v0.21}) as well as SDR objective metrics using libvmaf library  (\texttt{v2.3.1}).

\vspace{-1em}
\subsection{Participants and Data Processing}

\label{sec:sub-expt:data-process}
Currently, the ITU-R BT.500-14 (2019)~\cite{itubt50014} protocol recommends the minimum subjects to be 15. In our study, we recruited a total of 42 subjects (32 non-experts and 10 experts) from Trinity College Dublin (staff and students) with an average age is 29 (22-55). There were 30 males and 12 females. Consent forms were signed and data were anonymized. Following the recommendations of the BT.500-14 protocol, we performed outlier rejection on the collected scores. No participants were rejected. Moreover, we deployed {\em sureal} to account for subject bias and subject inconsistencies as per ITU-T Rec. P.910 (2022) [MOS-P910] and ITU-T Rec. P.913 (2021) [MOS-P913]~\footnote{https://github.com/Netflix/sureal, Access June. 2023}. 

\vspace{-0.5em}
\section{Results and Discussion}
\label{sec:result-discsn}


\noindent {\bf BD Measurements } Table~\ref{tab:obj-metrics} reports gains in terms of BD-Rate (\%) measured for the target 4 $qp$ points and for different objective quality metrics. Compared to the previous work, in this, we are reporting with 4 $qp$ points instead of 5 $qp$, and videos are of 5 seconds long instead of 130 Frames.
It is evident that, when we are optimising for a Luma based metrics (MS-SSIM), we observe considerable improvements (3.56 - 7.05\% on average) across all the Luma-based metrics, and we see a moderate degradation for Luma+Chroma-based metrics (up to 3.842\% for wPSNR-V on average) for some sequences.

\begin{figure}[t]
\captionsetup{font=small}

 \centering
  \begin{subfigure}[b]{0.32\columnwidth}
    \centering
    \includegraphics[width=\textwidth]{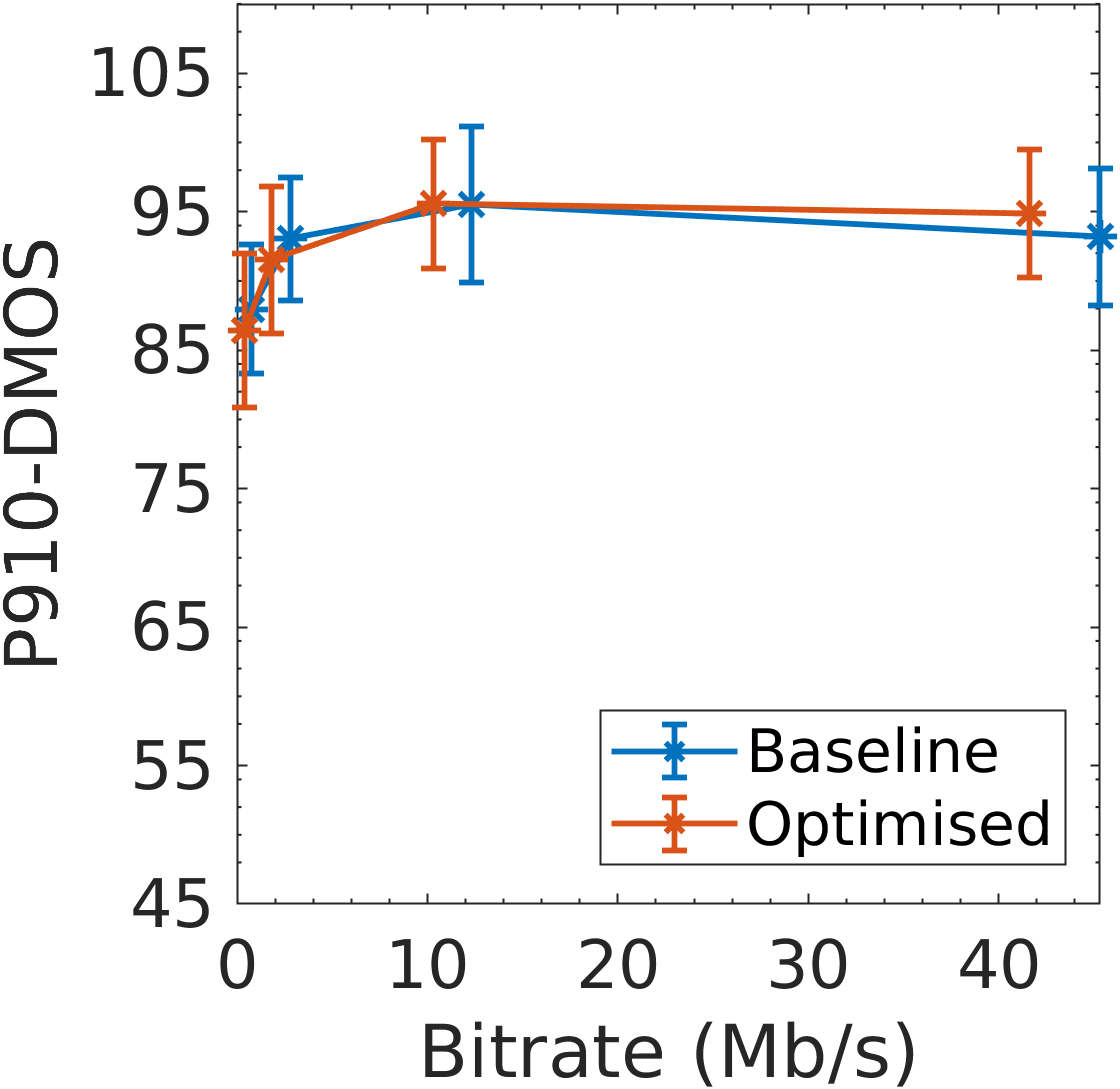}
    \caption{NocturneRoom}
    \label{fig:p913-mos-all-noctRoom}
  \end{subfigure}
  \begin{subfigure}[b]{0.32\columnwidth}
    \centering
    \includegraphics[width=\textwidth]{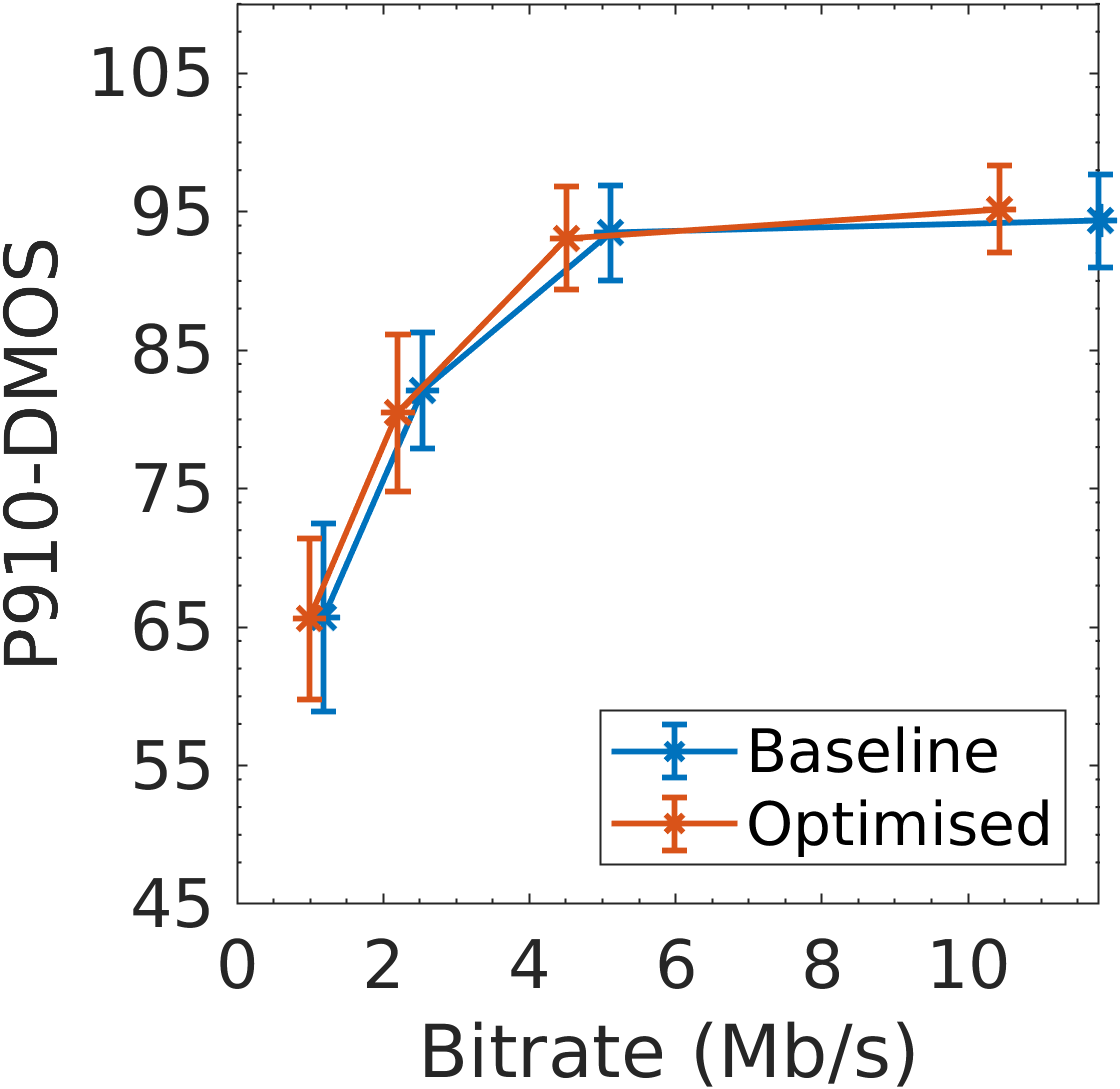}
    \caption{CosmosSheep}
    \label{fig:p913-mos-all-cosmos}
  \end{subfigure}
  \begin{subfigure}[b]{0.32\columnwidth}
    \centering
    \includegraphics[width=\textwidth]{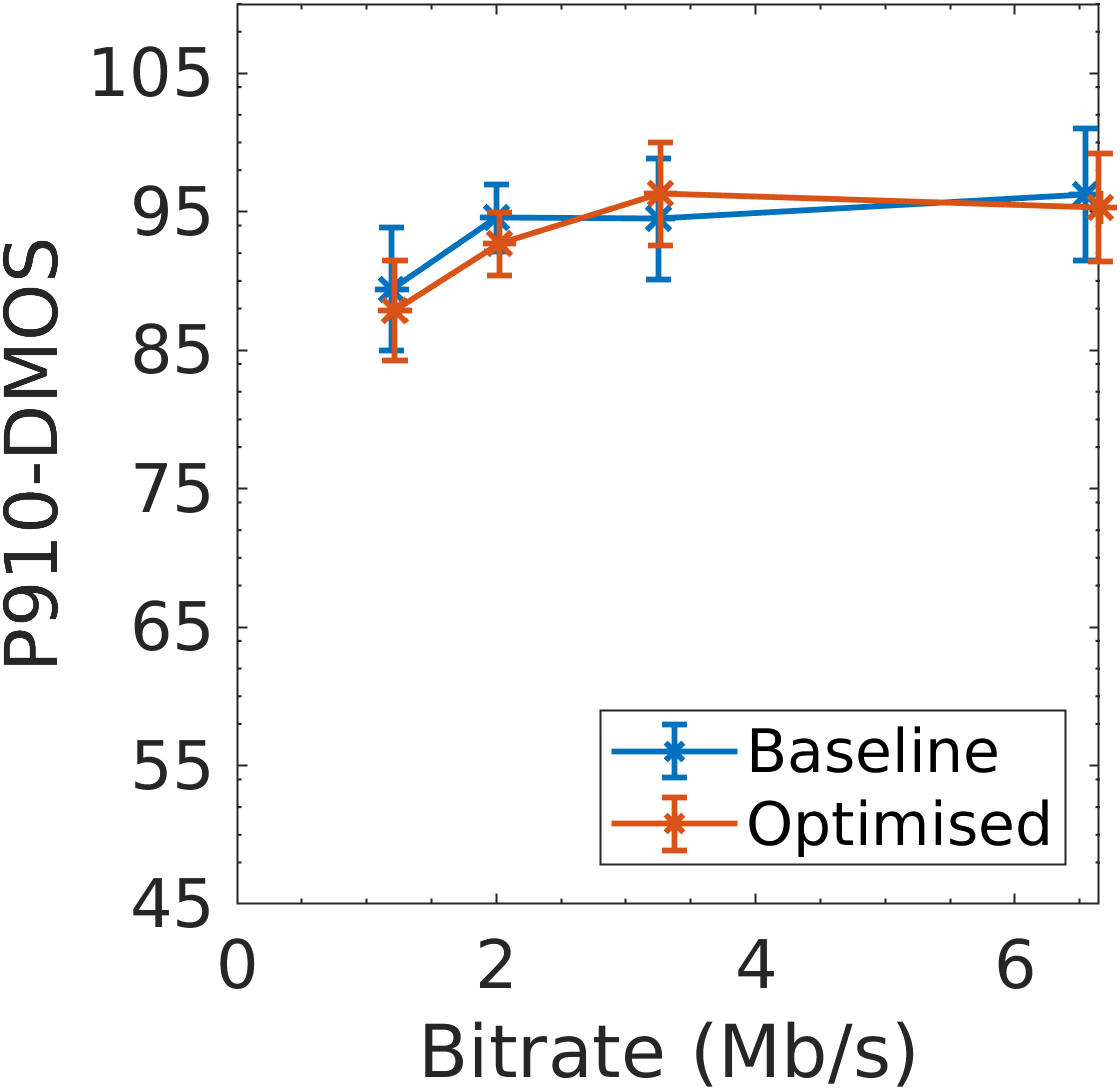}
    \caption{CablesButteryfly}
    \label{fig:p913-mos-wipe}
  \end{subfigure}
  \begin{subfigure}[b]{0.32\columnwidth}
    \centering
    \includegraphics[width=\textwidth]{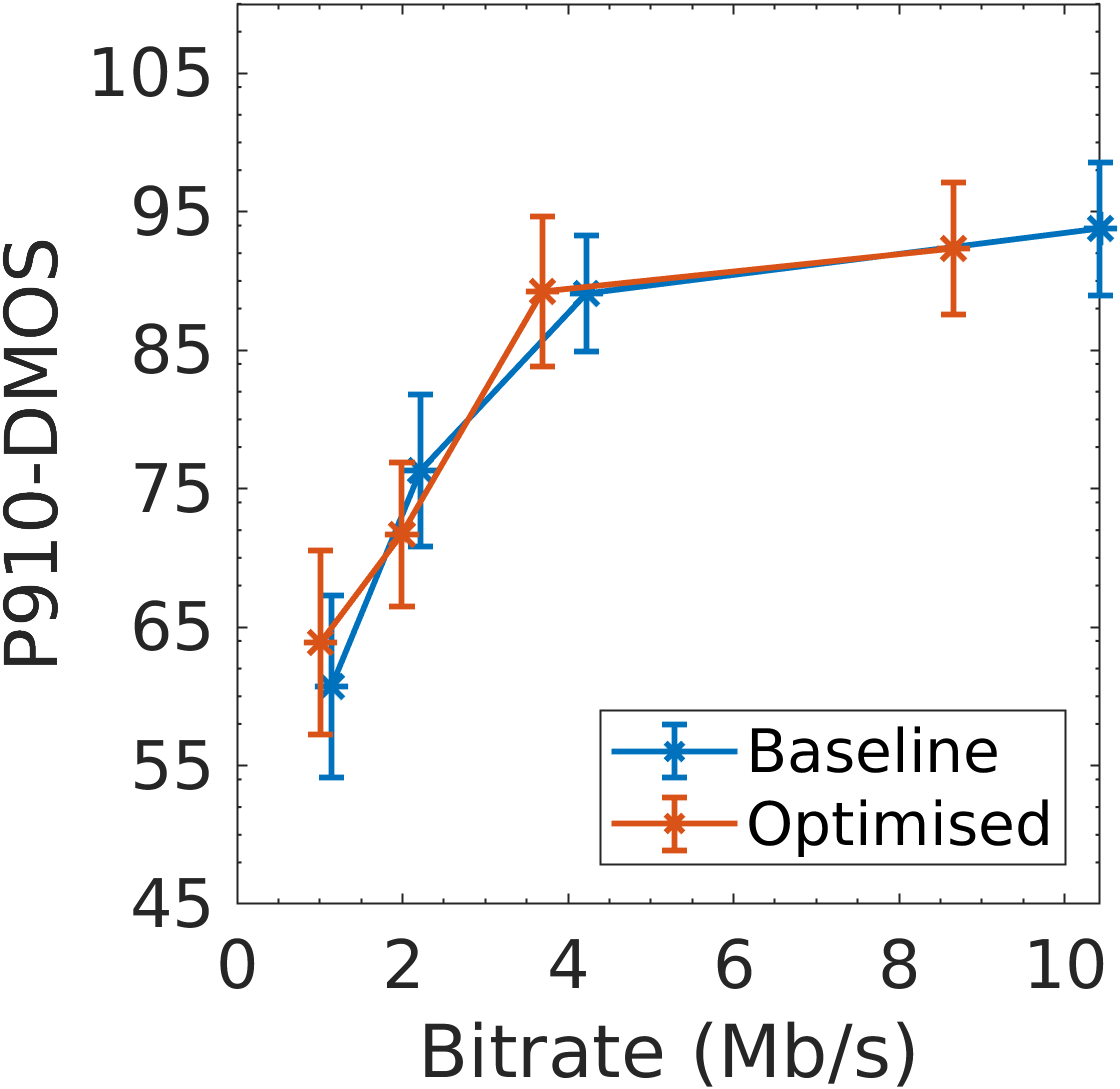}
    \caption{SLFireDragon}
    \label{fig:p913-mos-all-solevante}
  \end{subfigure}
  \begin{subfigure}[b]{0.32\columnwidth}
    \centering
    \includegraphics[width=\textwidth]{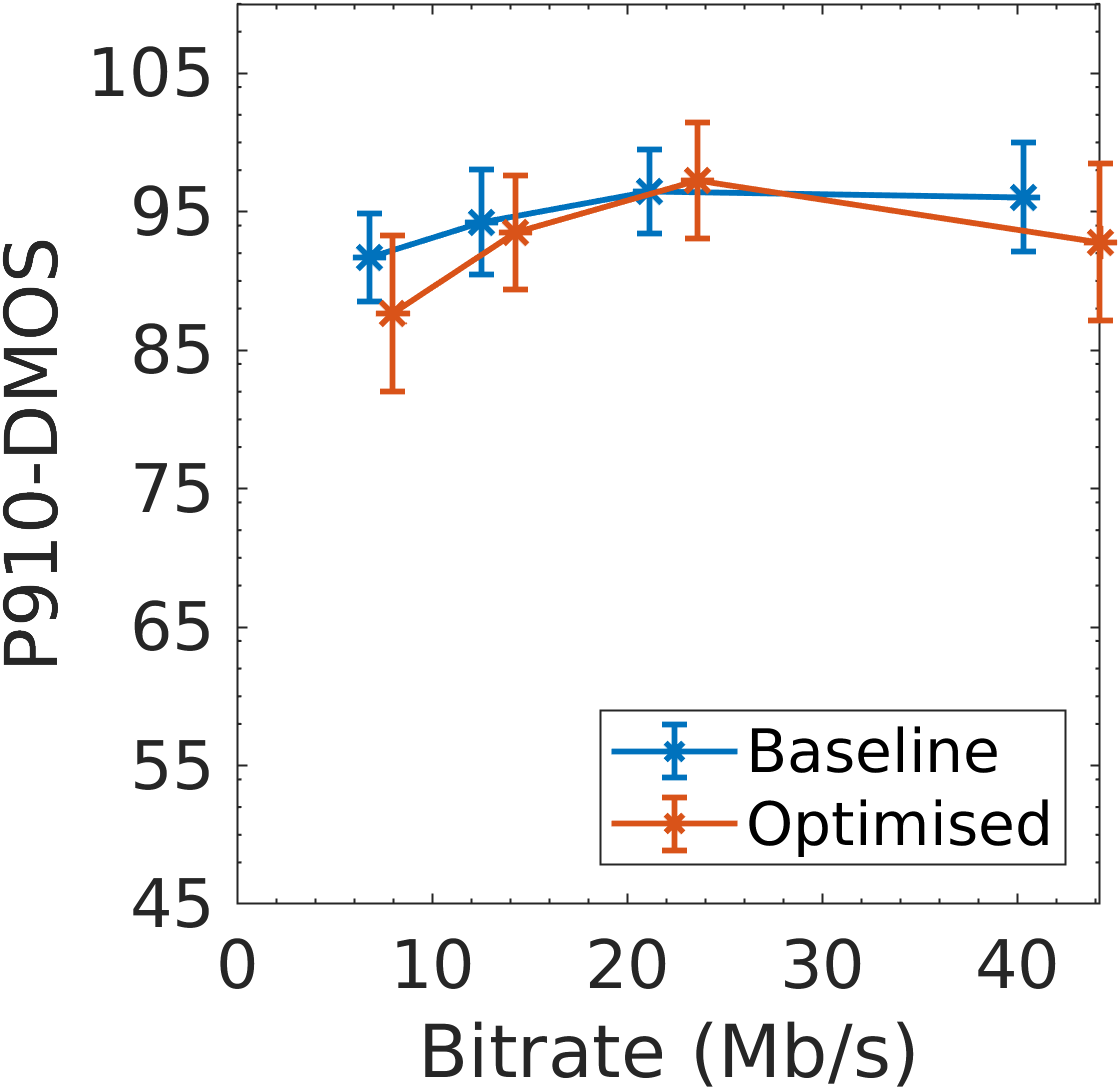}
    \caption{SVTSmithyHouse}
    \label{fig:p913-mos-svt}
  \end{subfigure}
  \begin{subfigure}[b]{0.32\columnwidth}
    \centering
    \includegraphics[width=\textwidth]{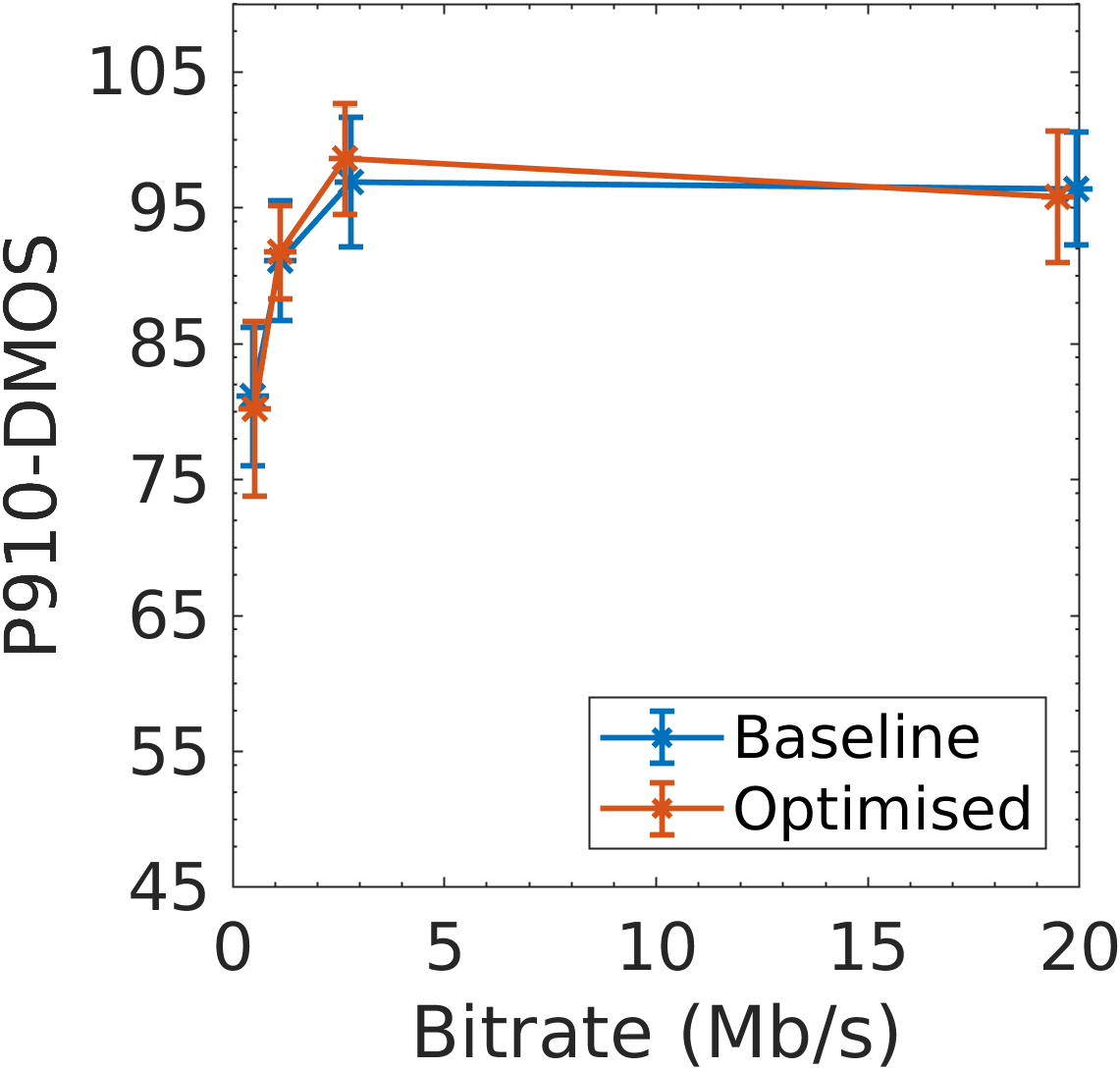}
    \caption{MeridianFace}
    \label{fig:p913-mos-all-merid}
  \end{subfigure}
  \begin{subfigure}[b]{0.32\columnwidth}
    \centering
    \includegraphics[width=\textwidth]{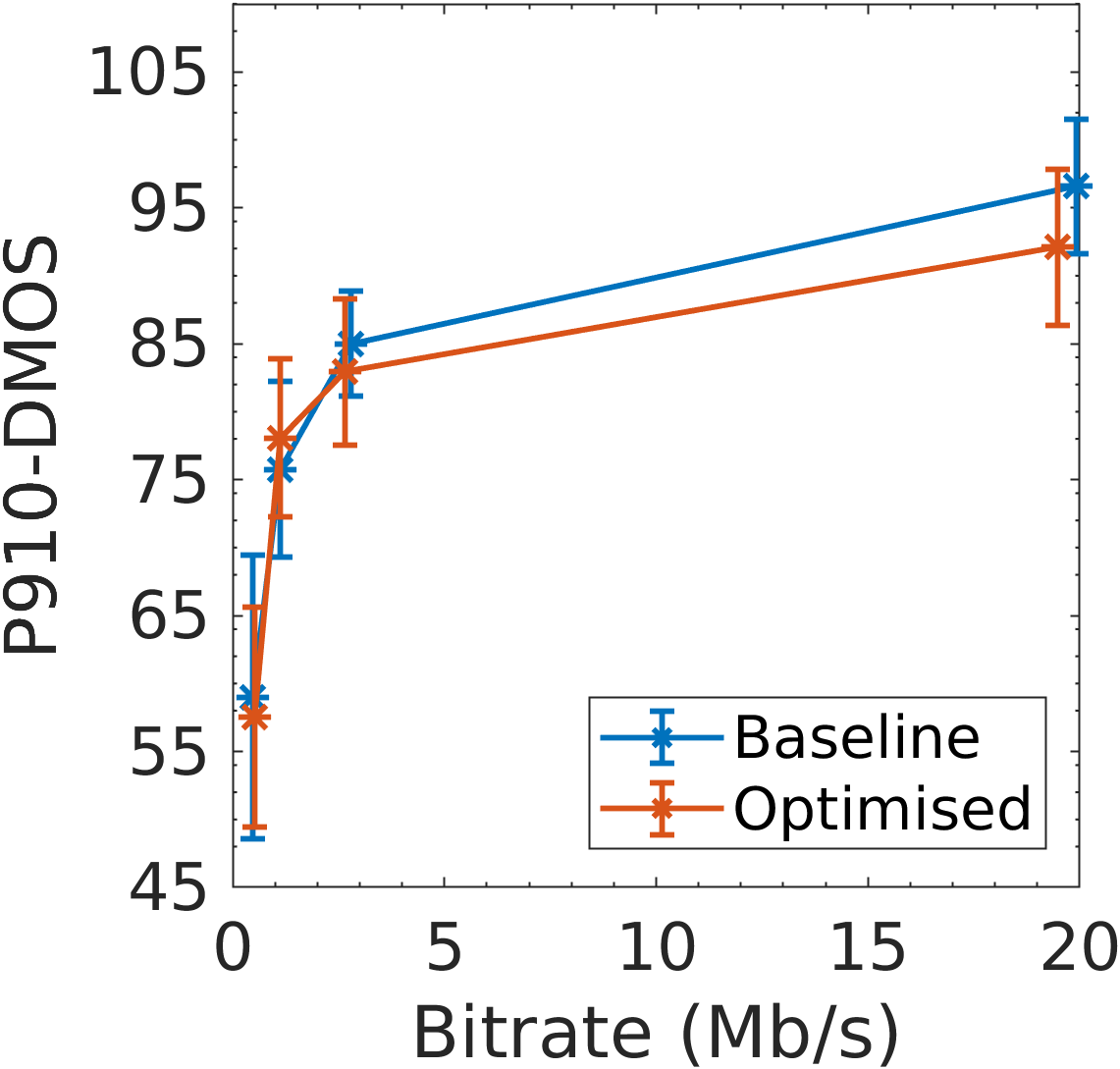}
    \caption{Experts-(f)}
    \label{fig:p913-mos-expt-merid}
  \end{subfigure}
  \begin{subfigure}[b]{0.32\columnwidth}
    \centering
    \includegraphics[width=\textwidth]{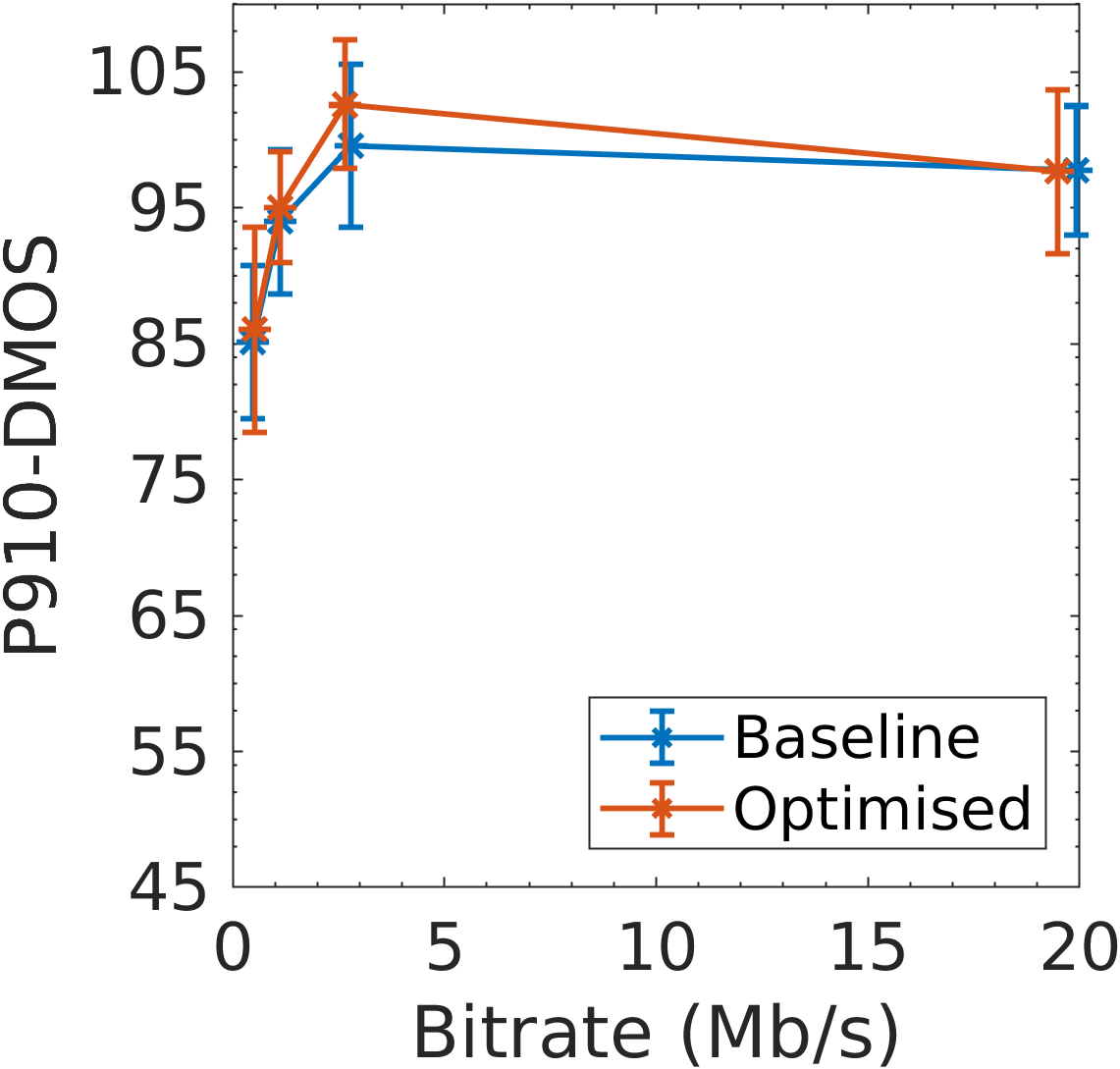}
    \caption{Non-Experts-(f)}
    \label{fig:p913-mos-noexpt-merid}
  \end{subfigure}
    \begin{subfigure}[b]{0.32\columnwidth}
    \centering
    \includegraphics[width=\textwidth]{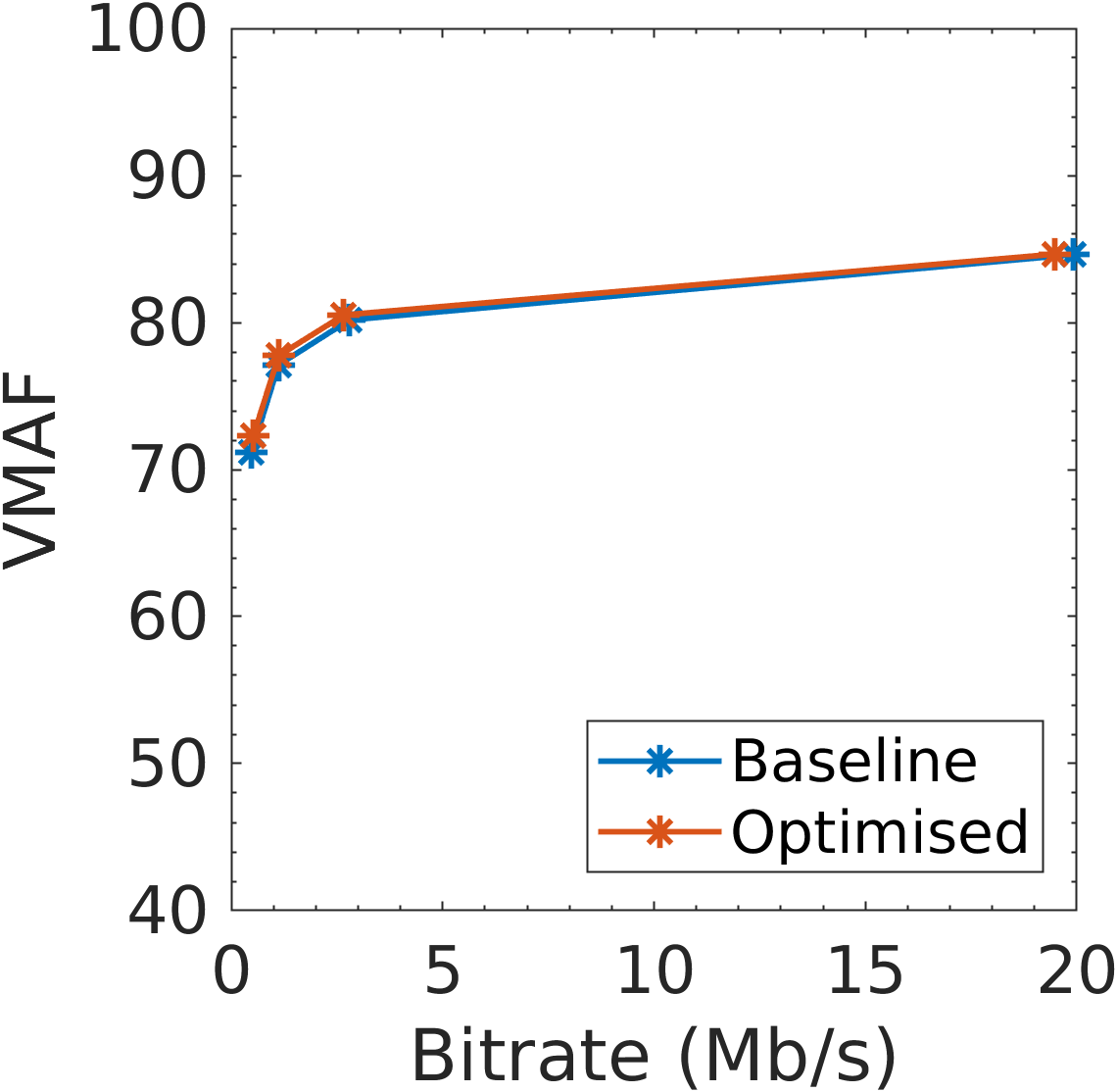}
    \caption{VMAF-(f)}
    \label{fig:p913-mos-vmaf}
  \end{subfigure}
  \vspace{-.5em}
  \caption{Fig~\ref{fig:p913-mos-all-noctRoom}-\ref{fig:p913-mos-all-merid}: P910-DMOS Score variation for all the 42 subjects who participated in the study, Fig~\ref{fig:p913-mos-expt-merid},~\ref{fig:p913-mos-noexpt-merid},~\ref{fig:p913-mos-vmaf} denotes P910-DMOS scores of MeridianFace sequence for a group of experts (10) and non-experts (32), VMAF respectively. }
    \vspace{-1em}
 
  \label{fig:mos-scores}
\end{figure}

\vspace{0.1cm}
\noindent {\bf Rate-Quality Curves~} Figure ~\ref{fig:mos-scores} (a)-(f) shows the Rate-Quality (RQ) curves for all the tested videos, here the distortion is represented as the Differential Mean Opinion Score \sloppy{($DMOS = 100-(MOS_{src} - MOS_{dist}$)} after recovery using the ITU-P910 method (42 subjects). The first observation from the curves is that they look very similar with overlapping confidence intervals across all tested content. We also notice that for the low-motion sequences the RQ curves become flat after the second compression level. The confidence ranges of MOS (DMOS), MOS-P910 (DMOS-P910) and MOS-P913 (DMOS-P910) are 3.50-7.55 (1.33-8.10), 2.79-5.73 (2.26-6.76), and 2.65-6.06 (2.29-6.84) respectively for all the 48 pairs.

\vspace{-0.6em}
\subsection{Data Analysis}


{\bf\noindent Optimised vs. Non-Optimised.} 
We aligned data points (using PCHIP interpolation) so that quality measurements aligned at the same bitrate. There are 2,016 data points (42 participants $\times$ 4 $qp$ levels $\times$ 6 videos $\times$ 2 optimisations). To examine the effect of optimisation on DMOS and MOS while controlling for $qp$, video and expert status of the participant, two four-way mixed-model analysis of variance (ANOVA) models were used with restricted maximum likelihood estimation \cite{Bates2015}. The models accounted for the within-subject nature of the experimental settings by including random effects for participants and participants by $qp$ interaction. The model included expert status as a fixed between-subject factor and $qp$, video and optimisation as fixed within-subject factors with all possible interactions of the fixed effects. $P$-values $<$ 0.05 from type II Wald F tests with Kenward-Roger degrees of freedom were considered significant \cite{Kenward1997}.

Considering two-way interactions for both MOS and DMOS, we find that video $\times$ expertise, video $\times$ $qp$ and expertise $\times$ $qp$ were significant with significant main effects. This means that the trends in RQ observed in these cases were significant. Furthermore, experts and non-experts differ significantly in their average MOS/DMOS scores but these differences depend on clips and $qp$. 

Neither the main effect for optimisation (i.e. the hypothesis that there is a change) nor any of the interactions involving the various possible dependent factors (e.g. video, qp, expertise) were significant w.r.t DMOS or MOS. 
Although the average quality improvement with optimisation is different, this statistical analysis shows that once all factors have been accounted for, there is no statistical significance in that observation at the $p$0.05 level. This could indicate that our subjective testing setup is not sensitive enough to detect the effect of optimisation, rather than prove that the effect does not exist.

{\bf\noindent Experts vs. Non-Experts.} 
We observed that for clips exhibiting ISO noise on the original, raw data (MeridianFace and NocturneRoom), experts consistently ranked $qp27$ higher than $qp39$ (preferring the higher bitrate encoding). In contrast, non-experts ranked $qp27$ lower than $qp39$ (preferring the lower bitrate encoding). Non-experts, therefore, rated these two clips completely differently from experts.

To assess the significance of this observation, a four-way mixed-model analysis of variance (ANOVA) model was used to test for the effect of the expertise factor and video clip on the change in DMOS at $qp27$ and $qp39$. The model was fitted separately for optimised and non-optimised versions of the videos because $qp$ levels were not equivalent (non-aligned bitrates). The models included the expertise factor of the participant, video and the two-way interaction as fixed effects. Random effects for participants were included to account for the within-participant repeated measures.  

The average difference in DMOS at $qp27$ and $qp39$ was higher for experts than non-experts in all 6 videos in both optimised and non-optimised versions. In both settings, the main effects were significant, but the interaction term was not, meaning that the effect of expertise on the change in DMOS was consistent across all videos. This shows that as the bitrate decreases, the experts' drop in DMOS was significantly higher than in non-experts. 
In optimised and non-optimised versions of MeridianFace and NocturneRoom (clips that exhibited ISO noise), the average DMOS difference for non-experts was negative, meaning that, on average,  a higher bitrate ($qp27$) was rated as lower quality. This effect was statistically significant in three of the clips: MeridianFace (optimised and non-optimised) and NocturneRoom (optimised) as 95\% confidence intervals based on standard errors of the estimates did not include zero. For the non-optimised version of NocturneRoom, the significance was borderline. 

We believe that non-experts, who are used to consuming compressed content on VoD services, consider ISO noise as a quality degrading factor. Since the video encoder acts as a denoiser~\cite{2022varounpreprocessor} at a mid-range of bitrates, non-experts would tend to rate these lower bitrate clips as higher quality. This was confirmed by discussions held with the participants after the end of the experiment. A similar phenomenon was observed in a subjective study with SDR videos for UHD transmission for HEVC in 2015~\cite{2015patrickuhdnoiseref} for sources with noise.


\begin{figure}[t]
\centering
\includegraphics[width=0.8\columnwidth]{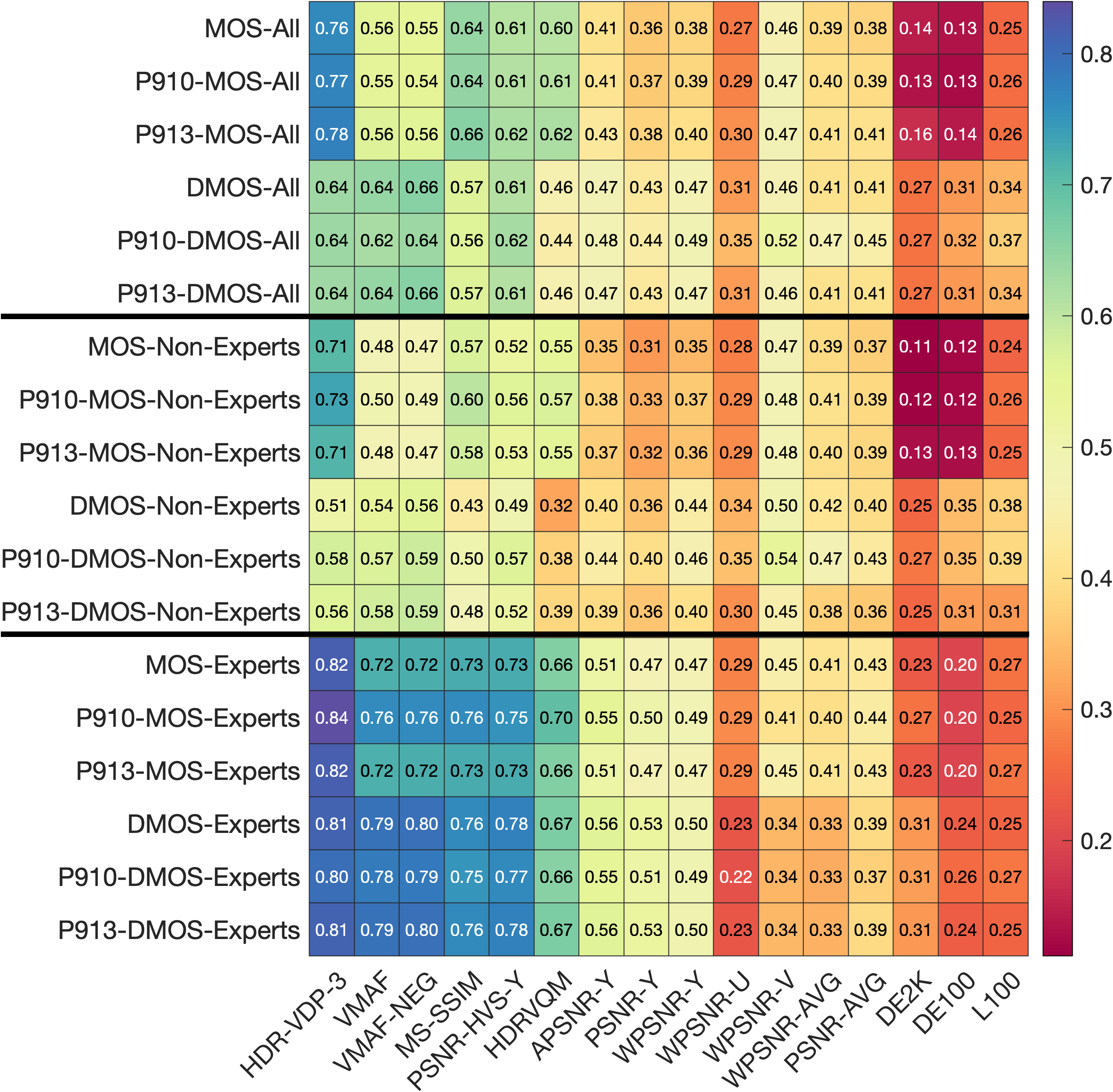}
\vspace{-0.2cm}
\caption{Spearman correlation for opinion scores from 42 subjects after non-linear mapping~\cite{2006subobjmapping} for the objective metrics after grouping by different cohorts and score recovery.}
\vspace{-1em}

\label{fig:spearman-corr-dmos}
\end{figure}

\vspace{0.1em}
{\bf\noindent Recommendations: }
As noted recently~\cite{2022reznikbdr}, computing BD-Rate for non-monotonic MOS-Rate curves with overlapping confidence intervals is problematic. 
We recommend that in this scenario where MOS scores are very similar, other subjective assessment protocols are more appropriate than DSCQS. In addition, non-monotonicity is not an indicator of a flaw in protocols but rather an expression of a fundamental truth about picture quality, that properties like film grain noise are not necessarily viewed as desired by experts and non-experts. The BD-RATE formulation as it exists today is not necessarily an appropriate measure in these cases unless the R-Q curves are pre-processed to take account of this trend.


\vspace{-0.5em}
\subsection{Correlation between objective metrics and subjective scores}
\label{sec:result-discsn:corr2-metr}
We mapped the subjective MOS scores to the computed objective metrics using a non-linear transform~\cite{2006subobjmapping} (a 5-parameter logistic function, constrained to be monotonic). 
This is used for comparing the objective metrics gains with all opinion scores. We computed Pearson, Spearman, and Kendall coefficients. Figure~\ref{fig:spearman-corr-dmos} shows the Spearman correlation between various objective metrics with subjective scores (MOS, DMOS) after recovery methods (P910/P913) separated by groups: all subjects, experts, and non-experts. Perceptual metrics like HDR-VDP-3, VMAF, MS-SSIM, PSNR-HVS-Y, and HDRVQM report a higher correlation to subjective scores consistently across different groups. The HDR-VDP-3 shows the highest Spearman correlation for P910-DMOS for different participant groups (All, Experts, Non-experts) as (0.64, 0.58, 0.80), followed by VMAF/VMAF-NEG (No-enhancement gain). While, within the objective metrics, HDR-VDP-3 is correlated with other perceptual metrics like MS-SSIM 
 (0.96) and HDR-VQM (0.93). This could be an indication that faster perceptual metrics even without HDR features (like MS-SSIM) may aid the per-clip optimisation process.  
Finally, we note that the experts MOS scores indicate a higher correlation to the objective metrics compared to the naive viewers, as anticipated from the analysis earlier.


\vspace{-1em}
\section{Conclusion}
\label{sec:conclusion}
In this work, we conducted subjective tests using the DSCQS protocol using 42 subjects targeting 4K HDR compression using a per-clip optimisation technique.
We observed gains of 5.19\% with 4.68\% bitrate savings on average (up to 14.78\%). Objective metrics in terms of SDR and HDR metrics have been analysed showing gains between 1.08\% and 6.08\%. Subjective analysis shows the perceived quality between experts and non-experts varies significantly over objective metrics when the video source has ISO noise. Due to the high variance in the subject responses, our experiments were not able to categorically say whether the proposed method yields perceivably better results than the one produced by the default encoder settings. 
Finally, among the objective quality metrics HDRVDP-3 and VMAF exhibit the highest correlation. The future work will include a study on what protocol yields minimum variance in subject results.

\bibliographystyle{IEEEbib}
\bibliography{refs}

\end{document}